\documentclass[aps,pra,twocolumn,showpacs,floatfix,superscriptaddress,10pt]{revtex4-1}

  \usepackage[utf8]{inputenc}
  \usepackage[T1]{fontenc}
  \usepackage[margin=2cm]{geometry}
  \usepackage{float}
  \usepackage{color}
\usepackage{graphicx}
\usepackage{amsmath}
\usepackage{amssymb}
\usepackage{bbm}
\usepackage{indentfirst}
\usepackage{dcolumn}
\usepackage{soul}
\usepackage{mathtools}

\usepackage{subfigure}
\usepackage{multirow}
\usepackage{setspace}

\begin{document}

\title{Metadynamics study of the temperature dependence of magnetic anisotropy and spin-reorientation transitions in ultrathin films}

\author{Balázs Nagyfalusi}
\email{nagyfalusi@phy.bme.hu}
\affiliation{Department of Theoretical Physics, Budapest University of Technology and Economics, Budafoki \'{u}t 8, H-1111 Budapest, Hungary}
\author{L\'{a}szl\'{o} Udvardi}
\author{L\'{a}szl\'{o} Szunyogh}
\affiliation{Department of Theoretical Physics, Budapest University of Technology and Economics, Budafoki \'{u}t 8, H-1111 Budapest, Hungary}
\affiliation{MTA-BME Condensed Matter Research Group, Budapest University of Technology and Economics, Budafoki \'{u}t 8, H-1111 Budapest, Hungary}
\date{\today}

\begin{abstract}
We employ metadynamics simulations to calculate the free energy landscape of thin ferromagnetic films and perform a systematic study of the temperature dependence of magnetic anisotropy and of the spin-reorientation transitions.  By using a simple spin model we recover the well-known power-law behavior of the magnetic anisotropy energy against magnetization and present a rather detailed analysis of the spin-reorientation transitions  
in ultrathin films. Based on tensorial exchange interactions and anisotropy parameters derived from first-principles calculations we perform simulations for Fe double layers deposited on Au(001) and W(110). In case of Fe$_2$W(110) our simulations display an out-of-plane to in-plane spin-reorientation transition in agreement with experiments. 
\end{abstract}

\maketitle

\section{introduction}

Since N\'eel's seminal paper in 1954 \cite{Neel1954} considerable interest has been focused on the magnetism of thin films
and multilayers. Magnetic anisotropy 
plays a key role in several phenomena important for technological applications.
In a magnetic data storage device the information is stored by controlling the magnetic orientation of a small magnetic domain that is
retained by magnetic anisotropy. In the early implementation of magnetic 
recording the magnetization of the bits were parallel with the plane of the film. Application of materials with perpendicular magnetic anisotropy
(PMA) triggered an order of magnitude increase of the storage density. The first realization of a 
perpendicular magnetic recording occurred  more than a decade ago \cite{PMR2005}, recent reviews on PMA can be found in Refs. \cite{PMA2017,
PMAoxid2017}. In order to further increase the storage density, the grain size in the recording medium should be decreased 
which requires a high uniaxial magnetic anisotropy of the thin film. Due to the large magnetic anisotropy of the recording media the field produced by the write head might no longer be sufficient to overcome the barrier to switch the 
magnetization. To circumvent this issue a heat assisted magnetic recording (HAMR) is proposed \cite{Hamann2004,McDaniel2005,HAMR2010}.
In HAMR the magnetic anisotropy is decreased by temporarily heating the domain storing the information. 

The temperature dependence
of magnetic anisotropy of thin films has been investigated both experimentally \cite{Okamoto2002,Thiele2002,Vaz2008,Fu2016}
and theoretically \cite{Staunton2004,Mryasov2005,Buruzs2007,Nowak2010}. The magnetic anisotropy energy (MAE) at finite temperature is usually defined as the 
difference between the free energy of the in-plane and that of the normal-to-plane magnetized system. 
Magnetic simulations provide different tools
for sampling the complex free-energy surfaces. One branch of such schemes 
is formed by the adaptive biasing potential methods  such as the Wang--Landau algorithm \cite{Wang2001},
umbrella sampling \cite{Marsili2006} and metadynamics \cite{Laio2002}.  In metadynamics a biasing potential is constructed 
as a sum of Gaussians centered along the trajectory in the space of the collective variables \cite{Laio2002}. 
In well-tempered metadynamics the smooth convergence of the biasing potential is guaranteed by changing adaptively the height of the Gaussians \cite{Barducci2008}. This algorithm is proved to converge to the exact free energy \cite{Dama2014}.

In this work we perform a systematic study of the temperature dependence of magnetic anisotropy and spin-reorientation transitions (SRT) by using metadynamics. In Section II we outline the main features of metadynamics simulations with the aim at studying the free energy landscape of a thin ferromagnetic film. In Section III we first present a model study of the temperature dependence  of magnetic anisotropy and a rather detailed analysis of the SRT 
in ultrathin films. Based on tensorial exchange interactions and anisotropy parameters derived from first-principle calculations we then present simulations on Fe bilayers deposited on Au(001) and W(110) and, finally, we summarize our results.

\section{Details of the metadynamics simulations}

The magnetic properties of thin films of transition metals are often described by classical spin models \cite{Nowak2007}. 
In most part of this work we choose a simple Heisenberg model to describe the magnetic properties of an ultrathin films
with uniaxial anisotropy and anisotropic exchange interactions:
\begin{equation} \label{eq:H}
 H = -\frac{1}{2} \sum_{\langle i, j \rangle} \left (J \, \mathbf{s}_i\mathbf{s}_j - d \, s_{zi}s_{zj} \right) - \sum_i \lambda_i  s_{zi}^2 \,,
\end{equation}
where $\mathbf{s}_i$ is a unit vector representing the direction of the 
atomic magnetic moment at site $i$, only nearest neighbors are considered in the first sum on the right-hand side with isotropic exchange coupling $J$ and an anisotropic part $d$, while $\lambda_i$
are the uniaxial anisotropy constants. More complex spin models will be presented and used only in Sections III.C and D in context of Fe bilayers on Au(001) and W(110). 

The free energy is sampled along an appropriately chosen collective variable (CV) labeled by $\eta$. For our present study we chose the $z$ (normal-to-plane) component of the normalized magnetization, 
$\eta = M_z/M$, as the collective variable, where $M_z = \sum_i s_{zi}$
and $M = \vert \sum_i \mathbf{s}_i \vert$. The key quantity in metadynamics is the bias potential $V_\mathrm{b}(\eta)$ added to the energy of the system.
Although in most of its applications metadynamics is implemented in molecular dynamics, there are examples where
it is successfully used  in Monte Carlo simulations \cite{Marini2008,Crespo2010} as well. If a Monte Carlo step (MCS) is interpreted as a time step the bias potential will be time dependent as well. 
After every $\tau$ MCS a Gaussan potential centered at the actual value of the CV, $\eta_{\mathrm{act}}$, is added to the bias potential:
\begin{eqnarray}
 V_\mathrm{b}(\eta,t+\tau) &=& V_\mathrm{b}(\eta,t) + V_{\mathrm{G}}(\eta - \eta_{\mathrm{act}}) \\
 V_\mathrm{G}(\eta-\eta_{\mathrm{act}}) &=& w e^{-\frac{(\eta - \eta_{\mathrm{act}})^2}{2\sigma}} \,, \label{eq:Vg}
\end{eqnarray}
where $\sigma$ and $w$ are the width and the height of the Gaussian, respectively. In well-tempered metadynamics \cite{Barducci2008}
the height of the Gaussian is chosen to change with the time $\tau$.

In our metadynamics simulations we applied a simple Metropolis algorithm \cite{Metropolis1953} with the probability of a random change of the spin at site $i$  $\mathbf{s}_i\rightarrow \mathbf{s}_i^\prime $,
\begin{equation}\label{eq:metropolis_prob}
 P(\mathbf{s}_i\rightarrow \mathbf{s}_i^\prime) = \mbox{min} \left\{
 1,e^{-\beta \left[ E(\mathbf{s}^\prime) + V_\mathrm{b}(\eta(\mathbf{s}^\prime)) - E(\mathbf{s}) -  V_\mathrm{b}(\eta(\mathbf{s}))
   \right] } \right\} \,,
\end{equation} 
where $\beta$ is the inverse temperature and $E(\mathbf{s})$ is the energy of the spin configuration given by Eq.~\eqref{eq:H}.
After a predefined number of Monte Carlo steps the biasing potential is updated by adding
a Gaussian centered at the actual value of the CV with the height of
$w=w_0e^{-\frac{V_\mathrm{b}(\eta)}{k_\mathrm{B}T_\mathrm{m}}}$,
where $T_\mathrm{m}$ is an appropriately chosen temperature as it is explained in the procedure of well-tempered metadynamics \cite{Barducci2008}. 
 In equilibrium, i.e. when the bias potential becomes stationary, the free energy $F(T)$ of the system is identified with the negative of the bias potential, $F(T)=-V_\mathrm{b}(\eta(T))$, where $\eta(T)$ stands for the equilibrium value of the CV  \cite{Dama2014}.

The values of the CV chosen for our model must be within the interval $[-1,1]$ 
and the free energy has a discontinuity at the boundaries which can not be accurately reproduced 
by a sum of finite-width Gaussians as it is detailed in Refs. \cite{Laio2008,Crespo2010}. In order to
eliminate this problem, the procedure  proposed by Crespo \textit{et al.} \cite{Crespo2010} has been modified in the following manner. Whenever the bias potential is updated, an extra Gaussian with the same width and height is added out of the physically relevant interval of the CV:
\begin{eqnarray}\label{eq:v_bias_mirror}
 V_\mathrm{b}(\eta,t+\tau) &=& V_\mathrm{b}(\eta,t) + V_\mathrm{G}(\eta - \eta_\mathrm{act}) \nonumber \\
                           &+& \left \{ \begin{array}{ll}                             
                                   V_\mathrm{G}(\eta - 2+\eta_\mathrm{act}) & \; \mbox{if} \; \eta > 0 \\
                                   V_\mathrm{G}(\eta + 2+\eta_\mathrm{act}) & \; \mbox{if}\; \eta < 0
                               \end{array}
 \right . \,,
\end{eqnarray}
where $V_\mathrm{G}(\eta)$ is the Gaussian potential given Eq.~\eqref{eq:Vg}. This scheme clearly makes the bias potential continuous at $\eta=\pm 1$. It should be noted that $V_\mathrm{b}(\eta)$ does not go smoothly to zero in the nonphysical region, but this part of the CV is never sampled during the simulation. In order to explore the free energy surface along the CV multiple walkers metadynamics \cite{Multiwalker}
was applied. The simulations were done simultaneously on typically four replicas  each contributing equally to the growth of a joint bias potential.

\section{Results and discussions}
\subsection{Temperature dependence of the magnetic anisotropy energy}
In order to validate metadynamics for the study of finite temperature magnetism
we first investigated the temperature dependence of the MAE, $K(T)$ of a monolayer. 
In case of on-site uniaxial anisotropy, the magnetic anisotropy energy should exhibit a $K(T)\propto M^3(T)$ scaling at low temperature as predicted by
Callen and Callen \cite{Callen1966}. If the magnetic anisotropy comes also from the exchange coupling, the low temperature behaviour of the MAE will be similar to the case of on-site uniaxial anisotropy, but at higher temperature the exponent will differ from three \cite{Staunton2004,Mryasov2005}.

The isotropic exchange couplings in ferromagnetic systems are closely related to the Curie temperatures. 
For ultrathin films of transition metals the Curie temperature is few hundreds of K and the corresponding effective
exchange coupling $J$  
is few tens of meV. 
The uniaxial anisotropy constant for hcp Co is 70\,$\mu$eV \cite{Baberschke2001}
and for a broad scale of thin films on different substrates it is in the range of 10-200\,$\mu$eV \cite{Johnson1996}.
According to the above experimental values, as compared to the effective isotropic coupling, the uniaxial anisotropy constant ($\lambda/J$) and the anisotropy of the exchange coupling  ($d/J$) have been chosen between $0.001$ and $0.01$ for the subsequent simulations.

\begin{figure}[htb!]
  \includegraphics[width=0.90\columnwidth]{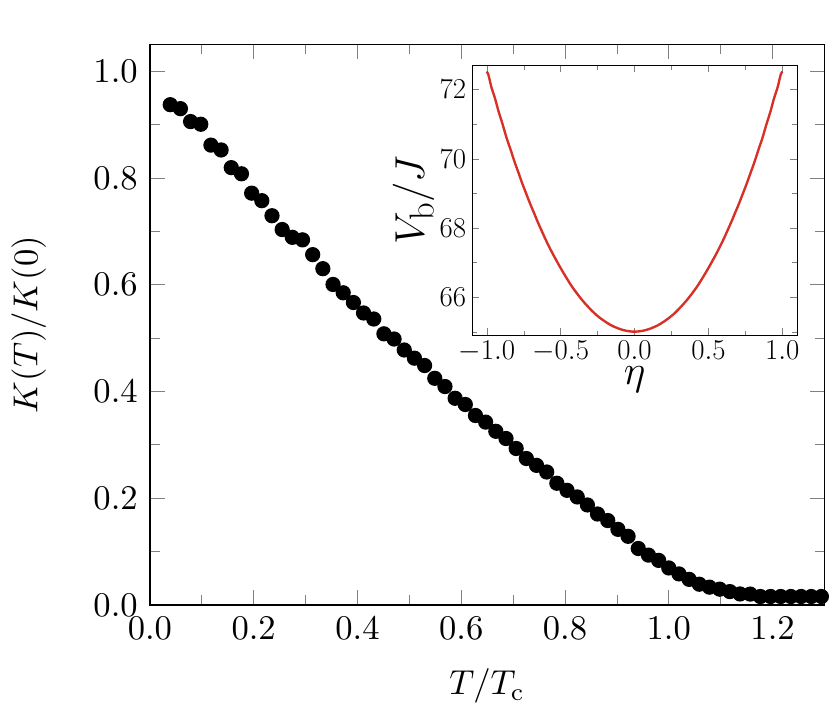}
 \caption{\label{fig:onsite-only} Temperature dependence of the magnetic anisotropy energy $K(T)$ of a monolayer with uniaxial on-site anisotropy $\lambda= 0.01\,J$ and vanishing exchange anisotropy $d=0$. In the inset the bias potential is shown at $T/T_\mathrm{C}=0.2$ as a function of the CV $\eta$. The simulations have been done on a $32\times32$ lattice with the parameters $T_\mathrm{m}=10\,J$, 
$w_0=0.02$ and $\sigma=0.03$ (see Section II). 
 }
\end{figure}

The first simulation  was  performed for a monolayer containing ferromagnetic nearest neighbour exchange coupling 
and uniaxial on-site anisotropy with easy axis perpendicular to the plane,
$(d=0,\;\lambda>0)$, see Eq.~\eqref{eq:H}. The ground state of the system is ferromagnetic with a normal-to-plane orientation. 
The bias potential has a quadratic dependence on the CV as it is  shown in the inset of Fig.~\ref{fig:onsite-only}. This parabolic 
behaviour is retained in the whole temperature range below the paramagnetic phase transition. The free energy has
a maximum at $\eta=0$ referring to the in-plane configuration and it has  minima at $\eta=\pm 1$ representing out-of-plane magnetic orientations. The difference between these two extrema 
is defined as the MAE.
Numerically more efficiently, $K(T)$ can be 
obtained as the second order coefficient of a symmetric parabola fitted to the bias potential. This is plotted in Fig.~\ref{fig:onsite-only}.  As the temperature
is increasing the curvature of the free energy (bias potential) as a function of CV is gradually decreasing and it
tends to zero above the Curie temperature. 
The Curie temperature is identified as the temperature 
corresponding to the maximum of the specific heat. Although the Curie temperature 
scales with the system size, it
should be chosen compatible with the size of the system for which the MAE
is calculated.

\begin{figure}[htb!]
  \includegraphics[width=0.90\columnwidth]{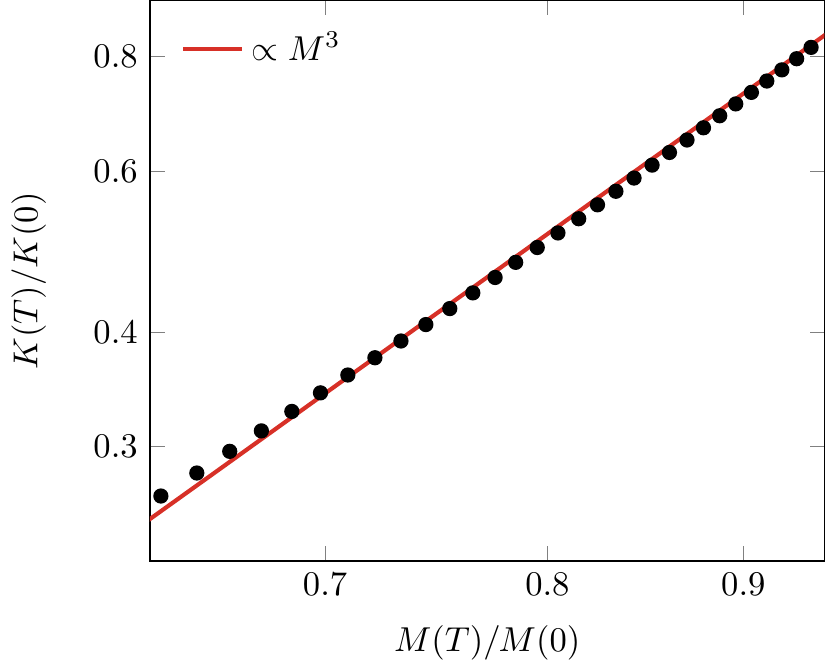}
 \caption{Log-log plot of the magnetic anisotropy energy of a $32\times 32$ square lattice with uniaxial anisotropy $\lambda=0.01\,J$ and vanishing exchange anisotropy $d=0$ as a function of 
the magnetization. Note that both the MAE and the magnetization are normalized to zero temperature. \label{fig:log_d0}}
\end{figure}

\begin{figure}[htb!]
  \includegraphics[width=0.90\columnwidth]{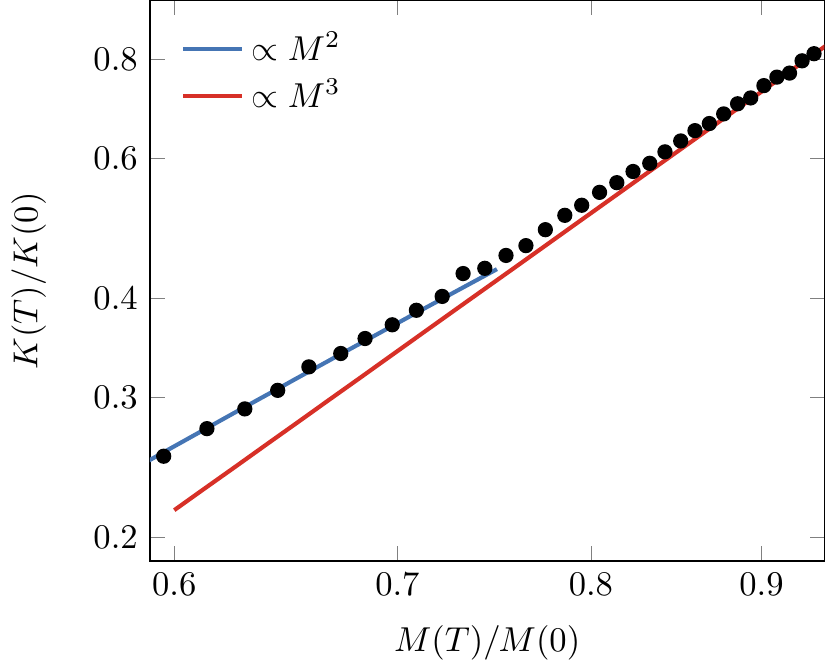}
 \caption{Log-log plot of the magnetic anisotropy energy of a $32\times 32$ square lattice with anisotropic exchange coupling $d=0.01\,J$ and zero on-site anisotropy $\lambda=0$ as a function 
 of the magnetization.  Note that both the MAE and the magnetization are normalized to zero temperature. \label{fig:log_lambda0}}
\end{figure}

The magnetic anisotropy in Fig.~\ref{fig:onsite-only} is almost linearly decreasing with the temperature similarly 
to the results obtained by using constrained Monte Carlo simulations \cite{Nowak2010} for uniaxial anisotropy.
The non-zero value of the magnetic anisotropy above the Curie temperature is the consequence of the finite size 
of the system. 
In Fig.~\ref{fig:log_d0} the MAE is plotted against the magnetization on a log-log mesh. As can be seen, at low temperatures
the results show excellent agreement with the scaling behavior predicted by Callen and Callen. 

If the uniaxial anisotropy $\lambda$ is removed from the model, Eq.~\eqref{eq:H}, and anisotropic exchange $d < 0$ is introduced, the scaling behaviour
of the anisotropy energy will be different as shown in Fig.~\ref{fig:log_lambda0}. 
At low temperatures the system behaves as in the case of uniaxial on-site anisotropy, but at higher temperatures the exponent $\gamma$
in the relationship $K(T) \propto M(T)^\gamma$ changes from three to two. Such a behavior of the temperature dependence of the
MAE was explored in earlier experimental \cite{Okamoto2002} and theoretical studies \cite{Staunton2004,Mryasov2005,Deak2014} 
for FePt alloys.

\subsection{Spin-reorientation transitions}

The interplay of different type of anisotropies often leads to a reorientation of the magnetization direction. The 
temperature driven spin-reorientation transition in thin films 
is usually explained by the competition of the uniaxial 
on-site anisotropy and the shape anisotropy \cite{Pappas1990,Fruchart1997, Sellmann2000, Slezak2010}. For planar systems the shape anisotropy due to the magnetic dipolar interaction always prefers in-plane magnetization, while the on-site anisotropy of a magnetic overlayer frequently prefers a normal-to-plane 
orientation. The shape anisotropy due to the anisotropic exchange interaction which is the consequence of the spin-orbit coupling 
may also prefer both directions.

\begin{figure}[h]
\includegraphics[width=0.90\columnwidth]{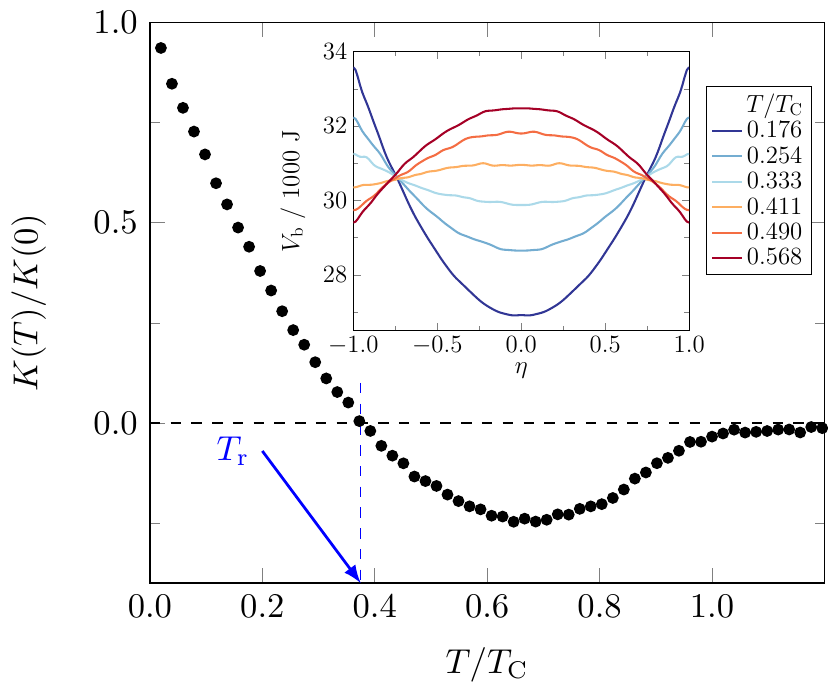}
 \caption{Magnetic anisotropy energy as normalized to zero temperature of a model monolayer system as a function of the temperature. 
 The metadynamics simulations have been done on 
a $64\times64$ rectangular lattice with competing on-site and nearest neighbor two-site anisotropy, $\lambda=0.05375\,J$ and $d=0.025\,J$, respectively. 
In the inset the bias potentials are shown for different temperatures as a function of the collective variable, 
$\eta=M_\mathrm{z}/M$. 
 }
  \label{fig:fordul_monolayer}
\end{figure}

In the model given in Eq.~\eqref{eq:H} the two competing anisotropies are the on-site uniaxial anisotropies $\lambda_i$ and the
anisotropy of the exchange coupling $d$. Considering a single square lattice,
in case of $\lambda > 2d$ 
the ground state is a normal-to-plane ferromagnetic. If $\lambda - 2d$ is not too large, a temperature induced normal-to-plane to in-plane SRT
can occur. In the inset of Fig. \ref{fig:fordul_monolayer} the bias potentials for a monolayer with $\lambda=0.05375 \, J$ and $d=0.025 \, J$ are 
shown for different temperatures as obtained from metadynamics simulations. At low temperatures the maxima of the bias potentials --- the minima of the free energy --- correspond to $\eta=M_\mathrm{z}=\pm 1$, i.e. to a 
normal-to-plane configuration. As the temperature is increasing the curvature of the bias potential changes sign and the minimum of the free 
energy moves to $\eta=M_\mathrm{z}=0$, i.e. to in-plane magnetic orientation. The magnetic anisotropy energy $K(T)$
in Fig.~\ref{fig:fordul_monolayer} is zero at the transition temperature $T_\mathrm{r}$.  It is worthwhile to mention that if 
the magnetization turns into the plane the system will have a gap-less magnetic excitation spectrum and long range magnetic 
order will no longer exist according to the Mermin-Wagner theorem. However, the magnetic anisotropy energy can still be  defined as the free-energy difference between the normal-to-plane and in-plane magnetic orientations.
The bias potential $V_{\text b}$ shown in the inset of Fig. \ref{fig:fordul_monolayer} demonstrates 
a first order phase transition. Moschel and Usadel \cite{Moschel1995} using MC simulations and Fridman 
{\it et al.} \cite{Fridman2002} applying a Hubbard-operator technique also confirmed that  a monolayer exhibits first order SRT.

In a case of a bilayer our simple model results in a more feature-rich
phase diagram where both first order and second order SRT 
can occur. A mean-field analysis of
a very similar model has been performed almost two decades ago \cite{Udvardi2001} and here we recall some of the results of this study. 
As a model system we consider a bilayer on an fcc(001) surface with nearest neighbor interactions $J$ and $d$, and on-site anisotropy parameters $\lambda_1$ and $\lambda_2$. 
At zero temperature supposing uniform magnetization within each monolayer, but different orientations in the two monolayers the energy of the system can be written as:
\begin{eqnarray}\label{eq:Ener} \nonumber
 E = &-& 4J + (2d-\lambda_1)\cos^2(\vartheta_1) + (2d-\lambda_2)\cos^2(\vartheta_2) \\
     &-& 4J\cos(\vartheta_1-\vartheta_2) + 4d\cos(\vartheta_1)\cos(\vartheta_2) \;,
\end{eqnarray}
 where $\vartheta_i$ is the polar angle with respect to the axis perpendicular to the layers ($z$). In the case of uniform in-plane
$(\vartheta_1=\vartheta_2=\pi/2)$ and a normal-to-plane $(\vartheta_1=\vartheta_2=0)$ orientations the energy has an
extremum. The energies of these two particular configurations coincide if $4d = \lambda_1 + \lambda_2$ defining a line in the
$\{\lambda_1,\lambda_2 \}$ parameter space. In the vicinity of this line a canted magnetic configuration exists. The boundaries
of the region of the canted states can be obtained from the stability condition:
\begin{equation}
 \left \vert \frac{\partial^2 E}{\partial \vartheta_i\vartheta_j} \right \vert_{\vartheta_i=0,\pi/2} = 0\;,
\end{equation}
yielding the lower boundary line,
\begin{equation}\label{eq:in-plane}
 \left(J + d - \frac{\lambda^\mathrm{l}_1}{2}\right ) \left(J + d - \frac{\lambda^\mathrm{l}_2}{2}\right ) = (d-J)^2
\end{equation}
and the upper boundary line,
\begin{equation}\label{eq:normal-to-plane}
 \left(J - 2d + \frac{\lambda^\mathrm{u}_1}{2}\right ) \left(J - 2d + \frac{\lambda^\mathrm{u}_2}{2}\right ) = J^2 \;.
\end{equation}
Below the line given by Eq.~\eqref{eq:in-plane}, $\lambda_1 + \lambda_2 <  \lambda^\mathrm{l}_1 + \lambda^\mathrm{l}_2$, the ground state is in-plane ferromagnetic 
and above the line given by Eq.~\eqref{eq:normal-to-plane}, $\lambda_1 + \lambda_2 >  \lambda^\mathrm{u}_1 + \lambda^\mathrm{u}_2$, it is normal-to-plane ferromagnetic.

At finite temperature the mean-field free energy of the double layer can be expressed as:
\begin{equation}
 F = \frac{4J}{2}\mathbf{M}^2  - \frac{4d}{2}{M^z}^2 - k_\mathrm{B}T\ln(Z_1) - k_\mathrm{B}T\ln(Z_2)
\end{equation}
where
\begin{eqnarray} \nonumber
 Z_i &=& \int I_0(4J\beta M^x\sin(\vartheta))
  \exp\left [4(J-d)\beta M^z \cos(\vartheta)\right ] \\
  & \times & \exp\left[\lambda_i\beta \cos^2(\vartheta) \right ] \sin(\vartheta)\mathrm{d}\vartheta \;,
\end{eqnarray}
$I_0(x)$ is the modified Bessel function of the first kind, while $M_x$ and $M_z$ are the $x$ and $z$ component of the 
magnetization of the bilayer, respectively. 
As was shown in Ref.~\cite{Udvardi2001} 
the magnetization can go to zero either via an in-plane or via a normal-to-plane direction at temperatures, 
$T_x$ and $T_z$, respectively, the higher of which can obviously be associated with the mean-field estimation of the Curie temperature $T_\mathrm{C}$.
Minimizing the free-energy with respect to the magnetization of the system with the constraint $M_z=0$ or $M_x=0$ 
and using a high temperature expansion yields the following expressions for $T_x$ and $T_z$ to first order in $\lambda_1$ and $\lambda_2$:
\begin{eqnarray}
 k_\mathrm{B}T_z &=&  \frac{8}{3}(J-d) + \frac{4}{30}(\lambda_1 + \lambda_2)\;, \\
 k_\mathrm{B}T_x &=& \frac{8}{3}J -  \frac{2}{30}(\lambda_1 + \lambda_2)\;.
\end{eqnarray}
An out-of-plane to in-plane SRT 
can occur only when the ground-state magnetization is out 
of plane and $T_z < T_x = T_\mathrm{C}$. In the case of a reversed SRT 
the ground state magnetization has to 
be in-plane (or canted) and $T_x < T_z = T_\mathrm{C}$. In the parameter space $\{\lambda_1,\lambda_2\}$
the region where SRT 
can occur are bounded by the line defined by Eq.~\eqref{eq:in-plane} and by the 
line where $T_x = T_z$: $\lambda_1 + \lambda_2 = \tfrac{40}{3}d$. 

\begin{figure}[htb!]
\includegraphics[width=0.99\columnwidth]{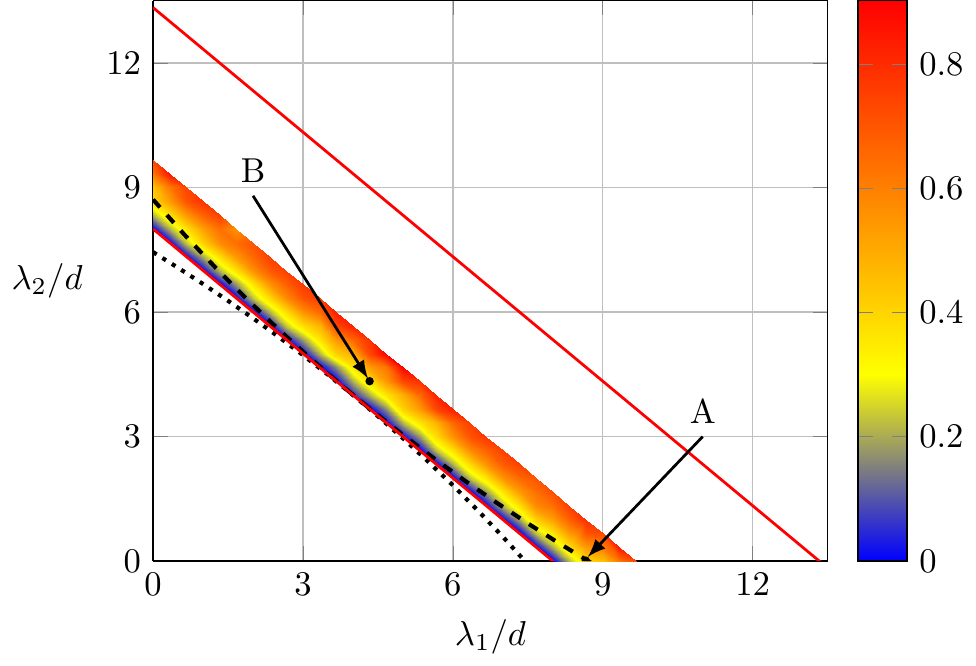}
 \caption{ Phase diagram of an fcc(001) ferromagnetic bilayer described by the model Hamiltonian Eq.~\eqref{eq:H} with nearest neighbor exchange interactions, $J$ and $d$, and uniaxial anisotropies, $\lambda_1$ and $\lambda_2$. For the case of $d=0.005\,J$, the lower solid red line shows the boundary where the normal-to-plane and in-plane configuration have the same energy, while the region of
canted ground states given by Eqs.~\eqref{eq:in-plane} and \eqref{eq:normal-to-plane} is comparable with the line width.
The upper solid red line bounds the area where a normal-to-plane to in-plane spin reorientation occurs according to mean-field theory. This area becomes considerably narrower from metadynamics simulations as shown by the colored area. 
The color-bar to the right refers to $T_\mathrm{r} / T_\mathrm{C}$.
  The dashed and dotted lines are the boundaries of the region with canted 
ground state for $d=0.05\,J$. For this case, the points A ($\lambda_1/d=8.66$, $\lambda_2/d=0$) and B ($\lambda_1/d=\lambda_2/d=4.33$) are chosen for further investigations, see text.}
  \label{fig:reorientationmap}
\end{figure}

We performed metadynamics Monte Carlo simulations to explore the phase diagram of a model bilayer. Although 
the anisotropy parameters $\lambda_i$ and $d$ can take both positive and negative values, in order to keep the MC simulations tractable, our investigations were restricted to the positive quarter of the parameter space $\{\lambda_1/d,\lambda_2/d\}$.
The phase diagram for $d=0.005\, J$ is shown in Fig.~\ref{fig:reorientationmap}.
In this case, the region where canted ground states exist   
determined by Eqs.~\eqref{eq:in-plane} and \eqref{eq:normal-to-plane} is extremely narrow. 
The area where a normal-to-plane to in-plane SRT 
 occurs provided by the metadynamics simulations (colored region) is considerably narrower then the corresponding area predicted by the mean field theory (bounded by the two solid red lines). 
The coloring clearly demonstrates that the
reorientation temperature $T_\mathrm{r}$ gradually approaches the Curie temperature as the uniaxial anisotropy constants are increasing, while parallel to the lines $\lambda_1+\lambda_2=\mathrm{const.}$ it is almost constant. If the uniaxial anisotropy is further increased the
system keeps its normal-to-plane ferromagnetic order till the ferromagnetic-paramagnetic phase transition. 

\begin{figure}[htb!]
\includegraphics[width=0.99\columnwidth]{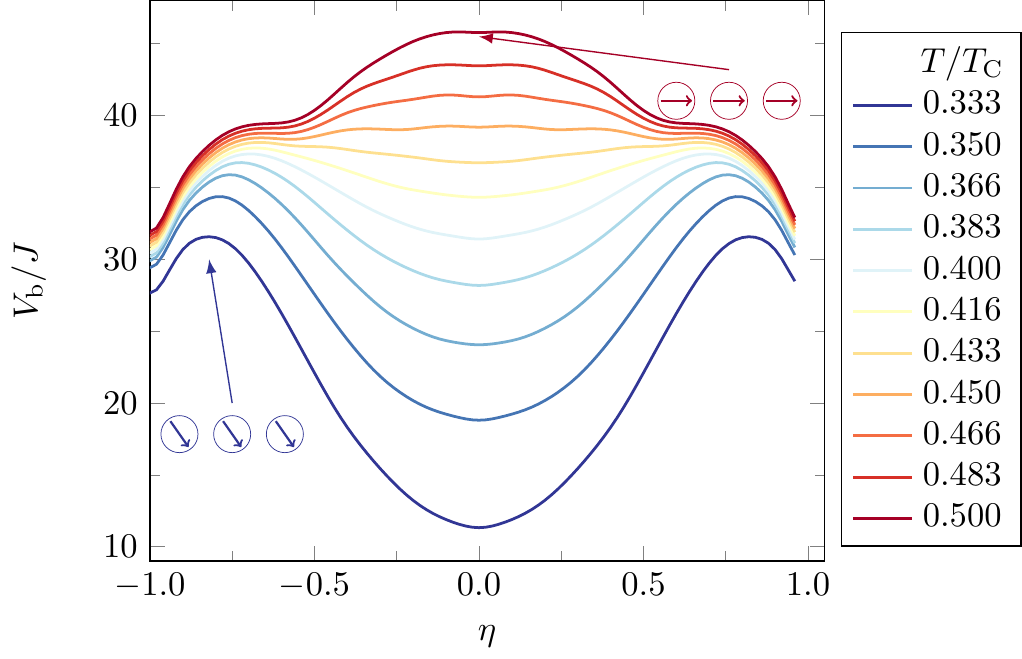}
 \caption{ Bias potentials from metadynamics simulations of a $2\times64\times64$ bilayer with nearest neighbor exchange interactions, $d=0.05\,J$, and anisotropy constants $\lambda_1=8.66\,d$, $\lambda_2=0$ (point A in Fig.~\ref{fig:reorientationmap}). The temperature is measured in units of $T_\mathrm{C}$. The low temperature magnetic configuration is canted ($0<\eta<1$), while by increasing the temperature the system continuously turns into the phase with in-plane magnetization showing the nature of a second order phase transition.
 }
  \label{fig:fordul-13}
\end{figure}

Increasing the two-site anisotropy $d$ the area of canted ground states on the phase diagram becomes wider. In the case of $d=0.05\,J$, the lower and upper boundary of the canted region are indicated by the dotted and dashed lines in Fig.~\ref{fig:reorientationmap}, respectively.
For further investigations we choose two points in the phase diagram: A ($\lambda_1/d=8.66$, $\lambda_2/d=0$) representing a canted ground state, however, lying in the vicinity of the upper boundary line of this region (dashed line in Fig.~\ref{fig:reorientationmap}) and B ($\lambda_1/d=\lambda_2/d=4.33$) corresponding to a normal-to-plane ferromagnetic ground state.
For the first choice of ($\lambda_1, \lambda_2$) 
the magnetization of the system continuously turns into the
plane as the temperature is increasing and, 
considering the normal-to-plane component of the magnetization as order parameter, 
the system undergoes a second order SRT. 
This is demonstrated in Fig.~\ref{fig:fordul-13}, where the bias potentials of a $2 \times 64\times 64$
lattice are shown as the function of the CV at different temperatures close to the SRT. 
Below the reorientation temperature, $T_\mathrm{r}/T_\mathrm{C} \sim 0.45$, the magnitude of the maximum position of the bias potential (minimum position of the free energy), $\eta_\mathrm{max}$, decreases continuously with increasing temperature, 
while at the in-plane magnetization $\eta=0$ there is a minimum in the bias potential. Above the reorientation transition temperature 
the in-plane configuration belongs to the maximum of the bias potential (minimum of the free energy), that means the order parameter is identical to zero.

\begin{figure}[htb!]
\includegraphics[width=0.99\columnwidth]{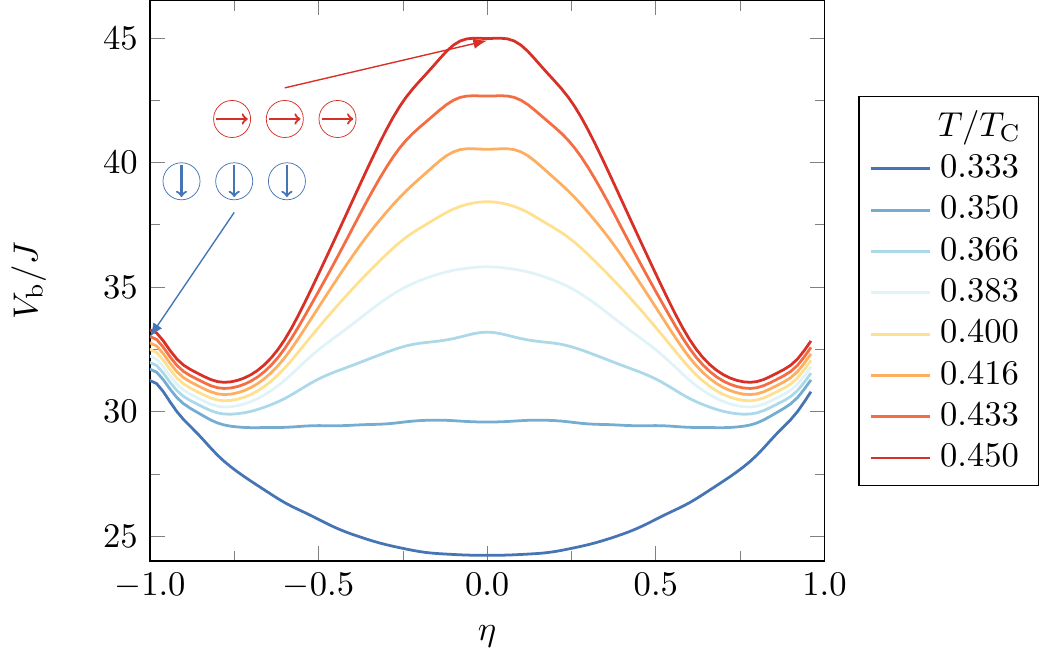}
 \caption{ Bias potentials from metadynamics simulations of a $2\times64\times64$ bilayer with nearest neighbor exchange interactions, $d=0.05\,J$, and anisotropy constants $\lambda_1=\lambda_2=4.33\,d$ (point B in Fig.~\ref{fig:reorientationmap}). The temperature is measured in units $T_\mathrm{C}$. At low temperature the magnetization points  normal to the plane ($\eta=\pm 1$), while at the reorientation temperature,
 $T_\mathrm{r} \sim 0.366 \, T_\mathrm{C}$, it suddenly jumps to in-plane, as $\eta=0$ becomes the maximum position of the bias potential (minimum position of the free energy), displaying thus a first order phase transition. }
  \label{fig:fordul_6p5_6p5}
\end{figure}

If the uniaxial anisotropy parameters are the same for both layers, $\lambda_1=\lambda_2$, no canted ground state exists for the bilayer, therefore, the mean-field description of temperature dependent magnetism is analogous with that of the monolayer.
The results of metadynamics simulations show, however, some different features
for the bilayer and the monolayer. According to Fig.~\ref{fig:fordul_monolayer} the SRT 
for the monolayer is discontinuous and the 
normal-to-plane and in-plane phases can not coexist. The bias potentials for the bilayer with anisotropy parameters $\lambda_1=\lambda_2=4.33\,d$ are shown in Fig.~\ref{fig:fordul_6p5_6p5}. Below the reorientation temperature the bias potential has maxima at $\eta=\pm 1$ which 
correspond to a normal-to-plane average magnetization. As the temperature is increasing a local maximum of the bias potential 
evolves at  $\eta=0$ referring to in-plane magnetization. Further increasing the temperature the local maximum at $\eta=0$
becomes the global maximum. The spin-reorientation transition is, therefore, of first order as in the case of the monolayer but the phases with in-plane and normal-to-plane magnetization can coexist.

\subsection{Fe$_2$Au(001)}

Over the past three decades, thin iron films deposited on the surface of gold have been the subject of extensive investigations, especially in context of low-dimensional magnetism, see e.g.\ Ref.~\cite{Wilgocka2010} and references therein. An Fe monolayer grown on Au (001) has often been referred as a prototypical two-dimensional ferromagnet. The film Fe$_n$Au(001) exhibits a normal-to-plane magnetic ground state for $n \leq 2$ and they undergo a 
thickness driven spin reorientation when the thickness of the Fe film reaches three monolayers \cite{Wilgocka2010}. While the driving force of this spin reorientation is the magnetostatic shape anisotropy, it is worth to study the temperature dependence of the spin-orbit induced MAE by using the metadynamics simulations introduced in this work. In this Section we present such a study for Fe$_2$Au(001).

For the simulations we used the following spin Hamiltonian,
 \begin{equation} \label{eq:Htensor}
	   H =- \frac{1}{2}\sum_{p,q=1}^2\sum_{i, j} \mathbf{s}_{pi}^T\mathcal{J}_{pi,qj}\mathbf{s}_{qj} - \sum_{p=1}^2\sum_i\lambda_p (\mathbf{s}_{pi}\hat{\mathbf{z}})^2 \, ,
 \end{equation}
 where $p$ and $q$ denote layers, $i$ and $j$ stand for Fe atoms within each layer, $\hat{\mathbf{z}}$ is a unit vector parallel to the $z$ axis, the $\mathcal{J}_{ij}$ is a $3\times 3$ matrix of exchange interactions and the sum in the first term is not restricted to the nearest neighbours only. The trace of the tensor $\mathcal{J}_{ij}$  can be identified as three times the isotropic
	exchange coupling $J_{ij}$, while the symmetric and anti-symmetric part of the tensor correspond to the pseudo-dipolar and
Dzyaloshinsky-Moriya interactions, respectively \cite{Udvardi2003}.
In order to determine the exchange tensors we applied the relativistic extension of the torque method \cite{Udvardi2003} 
 implemented in the framework of the Screened
Korringa-Kohn-Rostoker (SKKR) method \cite{skkr}.
 Since the (001) surface of fcc Au fits almost perfectly to the (001) surface of the bcc Fe (the lattice mismatch is less than 0.6\,\%) we used
two-dimensional translational symmetry for the whole system using the lattice constant of Au (2.87\,\AA). The Fe-Fe inter-layer distance has 
been chosen to be the same as the bulk value (1.44\,\AA) and the Fe-Au inter-layer distance was 1.6\,\AA. 

The calculated spin model parameters were then used in Monte Carlo and metadynamics simulations. 
In order to reduce finite size effects, the Curie temperature of the system has been determined from 
the intersection of the Binder cumulants yielding $T_\mathrm{C} \simeq 380\,\mathrm{K}$ in good agreement with
the experiments \cite{Durr1989}.
In order to characterize the anisotropy of the exchange tensors the lattice sum of 
the exchange couplings has been introduced:
\begin{equation}
 \mathcal{J}_p = \frac{1}{2}\sum_{q=1,2} \sum_j \mathcal{J}_{p0,qj} \;,
\end{equation}
where $\mathcal{J}_{p0,qj}$ is the coupling tensor between an arbitrary site $0$ in layer $p$ and site $j$ in the layer $q$.
Due to the C$_{4\mathrm{v}}$ symmetry of the lattice $\mathcal{J}_p$ is a diagonal matrix with identical $J_p^{xx}$ and 
$J_p^{yy}$ elements.

\begin{table}[ht!]
\caption{\label{tbl:FeAu} Calculated layer dependent magnetic anisotropy parameters (in units of meV) for the Fe$_2$Au(001) layers. The Fe layer at the interface with Au is denoted by I, the one at the surface by S. Negative/positive values of the anisotropies prefer  in-plane/normal-to-plane orientation of the magnetization.}
 \begin{tabular}{c|c|c} \hline
  layer     & $\lambda$ 		& $J_p^{zz} - J_p^{xx}$ \\\hline \hline
  I 	    & $-0.097$			& $\phantom{-}0.181$ \\
  S 	    & $\phantom{-}0.360$ 	& $-0.314$ \\ \hline
 \end{tabular}
\end{table}

\begin{figure}[ht!]
  \includegraphics[width=0.90\columnwidth]{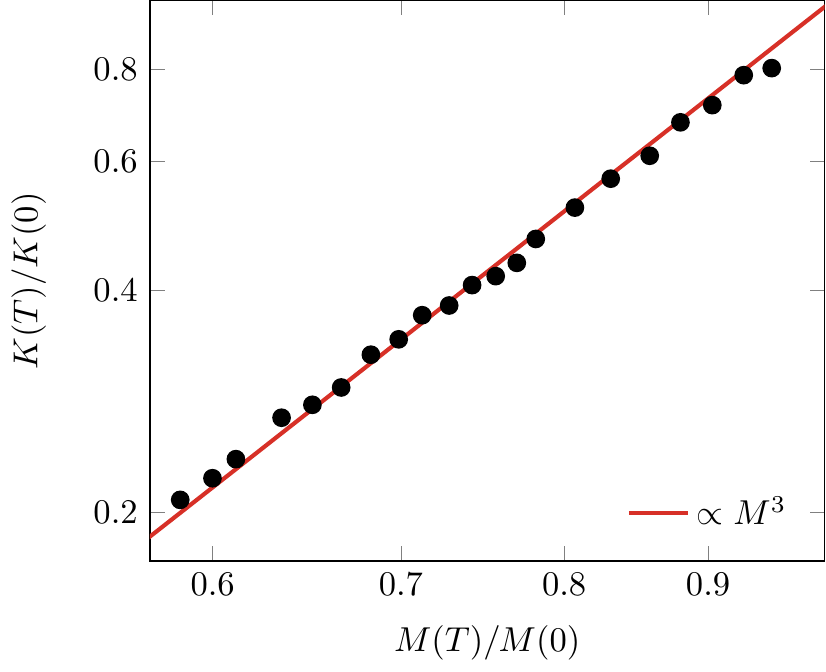}
 \caption{\label{fig:FeAuanis} Temperature dependence of the magnetic anisotropy energy of Fe$_2$Au(001).
The simulations were performed on a $2\times64\times 64$ lattice with meta temperature $T_\mathrm{m}=5\,J$, Gaussian height and width  are
 $w_0=0.08$, $\sigma=0.04$. }
\end{figure}

The layer dependent uniaxial anisotropy constants $\lambda_p$ and the anisotropy of exchange couplings $J_p^{zz}-J_p^{xx}$ 
are summarized in Table \ref{tbl:FeAu}. Interestingly, the on-site anisotropies and the exchange anisotropies have different signs in both the interface (I) and the surface (S) layer, and they also change sign between the two layers. Nevertheless, in both layers the positive contributions dominate, 
 resulting in an overall normal-to-plane magnetic ground state for the bilayer. The temperature dependent MAE obtained 
from metadynamics simulations is plotted in Fig.~\ref{fig:FeAuanis} at low temperatures as a function of the magnetization.
Similar to the model simulations, see Figs.~\ref{fig:log_d0} and \ref{fig:log_lambda0}, the MAE follows the regular $K \propto M^3$ rule \cite{Callen1966}.

\subsection{Fe$_2$W(110)}

Ultrathin Fe films epitaxially grown on W(110) have been studied intensively  
\cite{Sander1999,Meyerheim2001} due to their peculiar 
magnetic properties, such as in- and  out-of-plane anisotropy \cite{Hauschild1998}, 
spin  reorientation \cite{Weber1997,Slezak2010}, and domain wall formation  \cite{Heide2008}. 
In this Section we consider the double layer system Fe$_2$W(110). The magnetic ground state of this system strongly depends on the size 
and shape of the double-layer areas in the experiments  \cite{Bergmann2006,Slezak2010}. Fe DL stripes exhibit a
periodic magnetic structure with alternating out-of-plane domains separated by 180$^\circ$ walls \cite{Elmers1999}.
For larger DL islands there is a normal-to-plane ferromagnetic order at low temperature \cite{Weber1997}, which turns
into the $(1\overline{1}0)$ in-plane direction at higher temperature  
\cite{Dunlavy2004}.

As for Fe$_2$Au(001), the electronic structure of Fe$_2$W(110) was determined self-consistently via the SKKR method and the relativistic torque method was employed to find the exchange tensors and anisotropy parameters.
Since a DL of Fe grows pseudomorphically on W(110)\cite{Gradmann1982}, two-dimensional translational symmetry is applied
throughout the whole system with the lattice constant of bcc bulk W ($a_\mathrm{W}=3.16\,$\AA). According 
experimental \cite{Santos2016} and theoretical \cite{Qian1999} studies there is a considerable inward relaxation of the Fe layers due to the large lattice mismatch between Fe and W. Following Ref. \onlinecite{Heide2006}, the Fe-W and Fe-Fe layer distances were chosen as $2.01\,$\AA~and $1.71\,$\AA, respectively.  
In good agreement with previous calculations \cite{Qian1999,Heide2008}, we obtained $2.18\,\mu_\mathrm{B}$ and $2.73\,\mu_\mathrm{B}$ for the spin-magnetic moments of Fe in the surface and the interface layer,
respectively. 

We employed a spin model similar to that we used for 
Fe$_2$Au(001) layer, but because of the $C_\mathrm{2v}$ symmetry of the system bi-axial anisotropy applies,
 \begin{eqnarray} \label{eq:HFe2W}
	   H &=& - \frac{1}{2}\sum_{p,q=1}^2\sum_{i\neq j} \mathbf{s}_{pi}^T\mathbf{J}_{pi,qj}\mathbf{s}_{qj}  \nonumber \\
     & \ & + \sum_{p=1}^2 \sum_i\lambda_{px} (\mathbf{s}_{pi}\hat{\mathbf{x}})^2
      +  \sum_{p=1}^2 \sum_i\lambda_{py} (\mathbf{s}_{pi}\hat{\mathbf{y}})^2 \;, 
 \end{eqnarray}
where $\hat{\mathbf{x}}$ and $\hat{\mathbf{y}}$ are unit vectors parallel to the $(1\overline{1}0)$ and $(100)$ in-plane directions, respectively.
The layer-wise on-site and exchange anisotropy parameters, as explained in the case of Fe$_2$Au(001), are summarized in Table \ref{tbl:FeWanis}.
The anisotropy of the exchange couplings in the interface layer prefers the in-plane $(1\overline{1}0)$ direction which is
partially compensated by the contribution from the surface layer. On the contrary, the on-site anisotropy of the interface layer clearly prefers the $(110)$ direction for the magnetization. 
The MAE calculated as the difference between the energy of the system magnetized in the $(1\overline{1}0)$ in-plane direction and 
parallel to the normal-to-plane $(110)$ direction, $ E_{1\overline{1}0}-E_{110}=0.330\,\mathrm{meV}$,  as well as the MAE related to the $(001)$ and $(110)$ directions, 
$E_{001}-E_{110}=0.368\,\mathrm{meV}$, imply indeed a normal-to-plane magnetic orientation in the ground state as also found in Refs. \cite{Qian1999,Heide2008}.

\begin{table}[ht]
\caption{\label{tbl:FeWanis} Calculated layer dependent magnetic anisotropy parameters (in units of meV) for the Fe$_2$W(110) layers. The Fe layer at the interface with W is denoted by I, the one at the surface by S. The notations \textit{x},  \textit{y} and \textit{z} stand for the $(1\overline{1}0)$, $(001)$ and $(110)$ directions, respectively.}
 \begin{tabular}{c|r|r|r|r} \hline
  layer & \ \quad $\lambda_x$ \quad \ & \ \quad $\lambda_y$  \quad  \ & $J_p^{zz} - J_p^{xx}$ & $J_p^{zz} - J_p^{yy}$ \\\hline \hline
 I  &   0.611 $\:$ &  0.261 $\:$ & -- 0.603 $\;$ &  0.138 $\;$  \\
  S & $\:$ -- 0.055 $\:$ & $\:$ -- 0.137 $\:$ & 0.377  $\;$ &  0.106  $\;$  \\ \hline
 \end{tabular}
\end{table}

\begin{figure}[b!]
  \includegraphics[width=0.9\columnwidth]{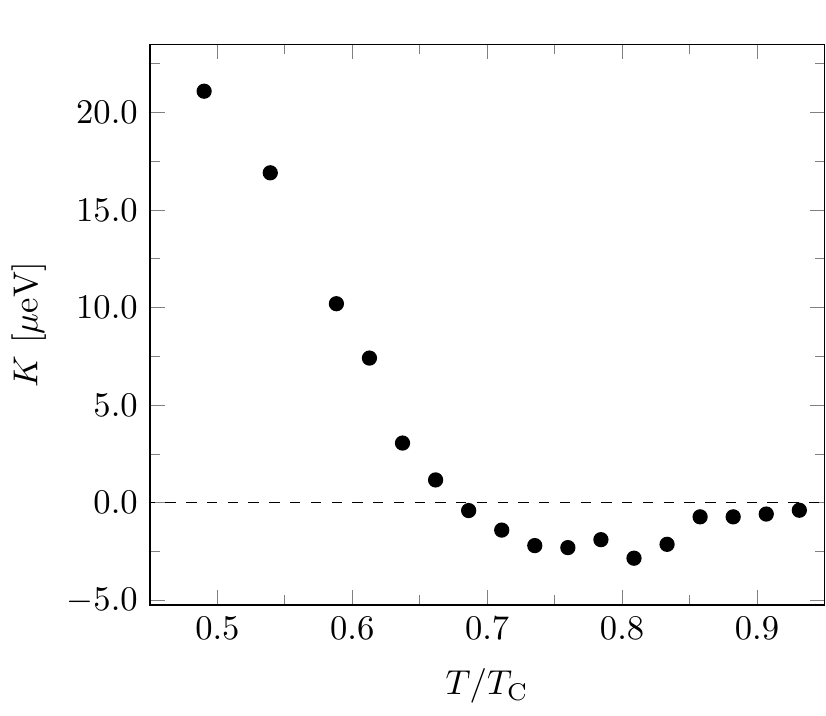}
 \caption{Calculated magnetic anisotropy energy, $K(T)=F_{1\overline{1}0}(T)-F_{110}(T)$, of an Fe DL on top of W(110) as obtained by well-tempered metadynamics MC simulations on a 
 $2\times64\times 64 \times$ lattice. The meta-temperature was chosen to be $k_{\rm B}T_\mathrm{m}=1\;\mathrm{mRyd}$, while
 the height and the width of the Gaussians were $w_0=0.06\;\mathrm{mRyd}$ and $\sigma=0.04$, respectively. 
 \label{fig:FeW}}
\end{figure}

According to susceptibility measurements \cite{Dunlavy2004} the Curie temperature strongly depends on the Fe coverage and in case of 1.8 monolayer of Fe  $T_\mathrm{C}=455\,K$ was measured. Our simulations on a perfect DL of Fe resulted in a Curie temperature of
520\,K, in relatively good agreement with the experiment.  In our metadynamics MC simulations the normal-to-plane component of the normalized magnetization was chosen again as the collective variable. In Fig. \ref{fig:FeW} the magnetic anisotropy energy defined as 
the difference of the free energy between the $(1\overline{1}0)$  in-plane orientation and the (110) normal-to-plane orientation
is depicted for a wide range below $T_\mathrm{C}$.   As can be inferred from this figure, the MAE changes sign at $T_{\rm r} =0.64\,T_\mathrm{C}$
indicating a SRT 
from the normal-to-plane to in-plane direction. 
The driving force of the spin reorientation is most probably a competition between the exchange anisotropy and the on-site anisotropy, since these contributions to the MAE exhibit different temperature dependence.

\section{Summary}

We introduced metadynamics as combined with Monte-Carlo simulations to study the thermal equilibrium of magnetic systems and demonstrated that the method can be applied to the temperature dependence of magnetic anisotropy of thin films. In particular, we  reproduced the power-law scaling of the magnetic anisotropy vs. magnetization proposed by Callen and Callen \cite{Callen1966} as for systems with on-site uniaxial anisotropy the simulations provided an exponent of three, whereas in case of dominating exchange anisotropy  
the exponent of two has been obtained in the high-temperature regime.
  
 We applied the method to explore spin-reorientation transitions in thin films. By using a simple spin model, first we performed a detailed analysis of the SRT for a monolayer and a double-layer. For double-layers we have shown that, by setting appropriate model parameters, both first and second order SRT can occur as it is predicted within the mean field theory. Then we considered two kinds of iron double-layer systems with perpendicular magnetic anisotropy where we set up a more complex spin-model containing tensorial exchange interactions calculated from first-principles methods. In case of Fe$_2$Au(001) the MAE followed the usual $M^3$ power law and no SRT was observed.  In case of Fe$_2$W(110) the MAE showed a more complex temperature dependence and our simulations reproduced the normal-to-plane to in-plane SRT seen in experiments \cite{Weber1997,Slezak2010}.  One of the future challenges for the simulations based on {\em ab initio} spin models is posed by exploring the effect of the Dzyaloshinsky-Moriya interactions on the temperature dependence of the magnetic anisotropy in thin films proposed recently \cite{Rozsa-DM-2017}.

 \bigskip
 \begin{acknowledgments}
The authors are grateful for the financial support by the Hungarian National Scientific Research Fund (NKFIH) under project No.\ K115575, as well as by the BME Nanotechnology FIKP grant of EMMI (BME FIKP-NAT). 
\end{acknowledgments}

\bibliography{metadynamics}{}
\end{document}